\documentclass[aps,twocolumn,showpacs]{revtex4}
\usepackage{graphicx}

\begin{document}

\title{Structures and Physical Properties of CsV$_2$Se$_{2-x}$O and V$_2$Se$_2$O}
\author{Hai Lin, Jin Si, Xiyu Zhu$^{*}$, Kehan Cai, Hao Li, Lu Kong, Xiaodong Yu, Hai-Hu Wen}\email{zhuxiyu@nju.edu.cn,hhwen@nju.edu.cn}

\affiliation{Center for Superconducting Physics and Materials,
National Laboratory of Solid State Microstructures and Department
of Physics, National Center of Microstructures and Quantum
Manipulation, Nanjing University, Nanjing 210093, China}

\date{\today}

\begin{abstract}
By using solid-state reactions, we successfully synthesize new oxyselenides CsV$_2$Se$_{2-x}$O (x = 0, 0.5). These compounds containing V$_2$O planar layers with a square lattice crystallize in the CeCr$_2$Si$_2$C structure with the space group of $P4/mmm$. Another new compound V$_2$Se$_2$O which crystallizes in space group $I4/mmm$ is fabricated by topochemical deintercalation of cesium from CsV$_2$Se$_2$O powder with iodine in tetrahydrofuran(THF). Resistivity measurements show a semiconducting behavior for CsV$_2$Se$_2$O, while a metallic behavior for CsV$_2$Se$_{1.5}$O, and an insulating feature for V$_2$Se$_2$O. A charge- or spin-density wave-like anomaly has been observed at 168 K for CsV$_2$Se$_2$O and 150 K for CsV$_2$Se$_{1.5}$O, respectively. And these anomalies are also confirmed by the magnetic susceptibility measurements. The resistivity in V$_2$Se$_2$O exhibits an anomalous log(1/$T$) temperature dependence, which is similar to the case in parent phase or very underdoped cuprates indicating the involvement of strong correlation. Magnetic susceptibility measurements show that the magnetic moment per V-site in V$_2$Se$_2$O is much larger than that of CsV$_2$Se$_{2-x}$O, which again suggests the correlation induced localization effect in the former.
\end{abstract}

 \maketitle

\section{Introduction}

The discoveries of cuprate\cite{LaBaCuO} and iron-based\cite{LaOFFeAs} high-temperature superconductors attract great attention on the investigation of 3$d$ transition-metal compounds. In these two systems, the CuO$_2$ and FeAs/FeSe planar layers with a square lattice play key roles in superconductivity. Scientists are inspired to seek more new superconductors from similar tetragonal layered 3$d$ transition-metal compounds. The $A$${T}$$_2$$Q$$_2$O compounds ($A$ = Na$_2$, Ba, (SrF)$_2$, (SmO)$_2$, (LaO)$_2$, etc.; ${T}$ = 3$d$ transition-meatal; $Q$ = S, Se, As, Sb, Bi, etc.) form a rich family which contains the [${T}$$_2$$Q$$_2$O]$^{2-}$ layers separated by the A$^{2+}$ units. The [${T}$$_2$$Q$$_2$O]$^{2-}$ layer consists of ${T}$$_2$O plane which has an anti-CuO$_2$ structure, and the two $Q$-anions locate just above and below the square plaque of $T$ atoms. Among the family, BaTi$_2$Sb$_2$O\cite{BaTiSbO}, Ba$_{1-x}$Na$_x$Ti$_2$Sb$_2$O\cite{BaNaTiSbO}and BaTi$_2$Bi$_2$O\cite{BaTiBiO} were discovered as superconductors with transition temperatures of $T_c$ = 1.2 K, 5.5 K and 4.6 K, respectively. The superconductivity seems to coexist with a charge- or spin-density wave (CDW/SDW) state. The (LaO)$_2$Co$_2$Se$_2$O\cite{LaOCoSe} and (LaO)$_2$Fe$_2$Se$_2$O\cite{LaOFeSe} were shown to be  anti-ferromagnetic(AFM) Mott-insulators, which are similar to the parent phase of cuprate superconductors. Recently, Le ${et al}$. suggested that the Ni-based compounds (LaO)$_2$Ni$_2$Se$_2$O containing the [Ni$_2$$M$$_2$O]$^{2-}$ ($M$=chalcogen) layers with the similar structure mentioned above may be potential high-temperature superconductors upon doping or applying pressure\cite{LaONiSe}. Furthermore, another system CsV$_2$S$_2$O\cite{CsVSO} with the isostructure but different-valent [V$_2$S$_2$O]$^{-}$ layer was reported to be a paramagnetic bad-metal. Therefore it is very interesting to explore and investigate the 3$d$ (probably also 4$d$) transition-metal systems with the similar [${T}$$_2$$Q$$_2$O] structural units.

In this paper, we report the synthesis and investigation on the new vandium oxyselenides CsV$_2$Se$_{2-x}$O(x = 0, 0.5) and V$_2$Se$_2$O. The powder X-ray diffraction (XRD) confirms that these compounds have the same [V$_2$Se$_2$O] layers, which belong to the family of $A$${T}$$_2$$Q$$_2$O. Through the measurements of electrical resistivity and magnetic susceptibility, we find that (i) The CsV$_2$Se$_{1.5}$O is a metal with an anomaly at about 150 K; (ii) The CsV$_2$Se$_2$O shows a semiconducting behavior with an anomaly at about 168 K; (iii) The V$_2$Se$_2$O is an insulator with the temperature dependence of $\rho$ $\propto$ log$(1/T)$ in wide temperature region, suggesting a Mott insulator behavior.

\begin{figure}
  \includegraphics[width=8.5cm]{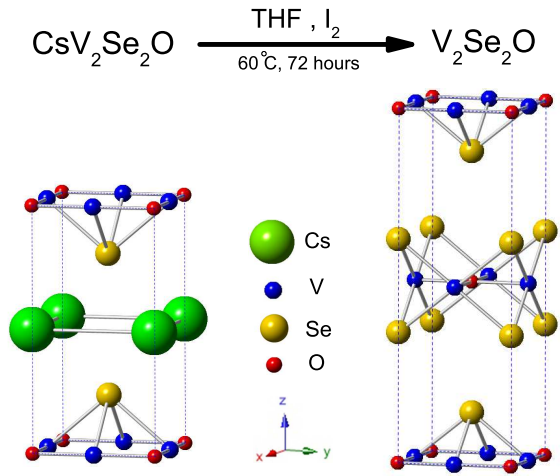}
\caption{Synthesis and Schematic structures of CsV$_2$Se$_2$O and V$_2$Se$_2$O. } \label{fig1}
\end{figure}

\section{Experimental details}

The CsV$_2$Se$_{2-x}$O(x = 0, 0.5) compounds investigated in this work were synthesized by solid state reactions. Firstly, the precursor VSe$_2$ was obtained from the reaction of V powder(99.9${\%}$ purity, Aladdin) and Se powder(99.999${\%}$ purity, Alpha Aesar) at 700$^{\circ}$C for 24 hours. Next, Cs pieces (99.9${\%}$ purity, Alpha Aesar), VSe$_2$ powder, V$_2$O$_5$ powder(99.5${\%}$ purity, Strem Chemicals) and V powder(99.9${\%}$ purity, Aladdin) were weighed according to the stoichiometric ratio, put into an alumina crucible and then sealed into two layer evacuated quartz tubes. These procedures should be performed in a glove-box filled with high-purity argon atmosphere. Then, the tubes were heated in the furnace at 1080 $^{\circ}$C for 15 hours. After the pre-reaction, the mixture was ground in an agate mortar, pressed into pellet and annealed at 1080 $^{\circ}$C for 10 hours, to make the reaction more uniform and perfect. Finally, we obtained the samples which are black and insensitive to air or water.

The V$_2$Se$_2$O was synthesized by topochemical deintercalation method. Firstly, the CsV$_2$Se$_2$O powders and I$_2$ grains(99.99${\%}$ purity, Alpha Aesar) were mixed in 1:1 molar ratio and put into a teflon-linked stainless-steel autoclave, which was filled with 10 mL water-free THF liquid. Then, the autoclave was sealed and heated up to 60 $^{\circ}$C and kept for 72 hours. The excess iodine reacted with Cs-ions generating CsI and extracts all of Cs from CsV$_2$Se$_2$O. The reaction process is displayed in the upper scheme of Figure~\ref{fig1}. Next, the product was leached by deionized water to wash out CsI and THF. Finally, powders in black color can be obtained after dried in nitrogen atmosphere at 60 $^{\circ}$C. As shown in Fig~\ref{fig1}, to keep the structure stable, every neighbouring V$_2$Se$_2$O layer should slide by 1/2 lattice constant along $a$- and $b$-axes. The similar method has been used in synthesizing Na$_{1-y}$Fe$_{2-x}$As$_2$ system, MoS$_2$, tetragonal Fe/Co chalcogenides and so on\cite{NaFeAs,MoS,LiMoS,FeSe,FeS,CoSeCoS}.

In this study, powder X-ray diffraction (XRD) data was obtained by using a Bruker D8 Advanced diffractometer with the Cu-K$_\alpha$ radiation. The Rietveld refinements\cite{Rietveld} are done with the TOPAS 4.2 software\cite{TOPAS}. The DC magnetization measurements were performed on a SQUID-VSM-7T (Quantum Design). The resistivity measurements were done on a Quantum Design instrument Physical Property Measurement System (PPMS) with the standard four-probe method.

\section{Results and discussion}
\subsection{Sample characterization}

\begin{figure}
  \includegraphics[width=8.5cm]{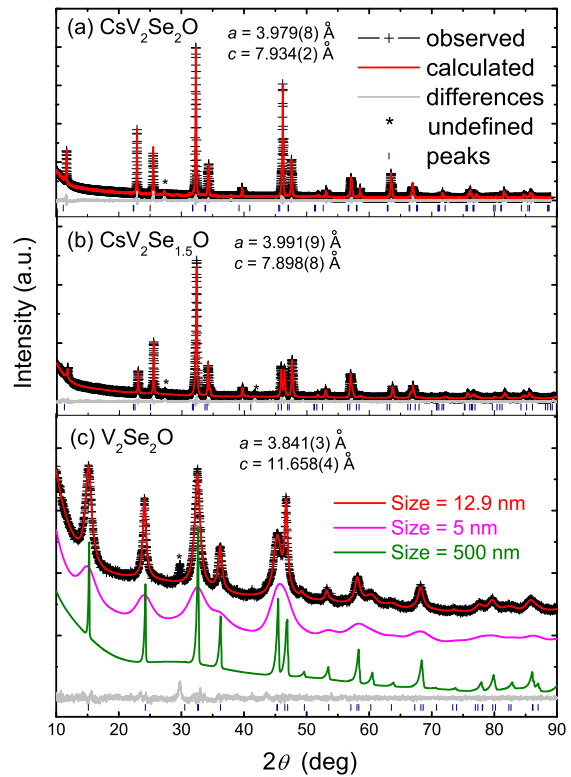}
\caption{Observed (black crosses) and calculated (red lines, using Rietveld method) powder X-ray diffraction data of the (a) CsV$_2$Se$_2$O, (b) CsV$_2$Se$_{1.5}$O and (c) V$_2$Se$_2$O polycrystalline samples, together with the differences (gray line) between them and the Bragg peak positions (vertical lines). The pink and green lines of (c) show the calculated results using different domain sizes.} \label{fig2}
\end{figure}

Figure~\ref{fig2} shows the XRD patterns of the (a) CsV$_2$Se$_2$O, (b) CsV$_2$Se$_{1.5}$O and (c) V$_2$Se$_2$O polycrystalline samples. They can be refined very well by Rietveld method. The room-temperature structural parameters obtained from the Rietveld refinement profiles are shown in Table~\ref{tbl}. The observed diffraction peaks of CsV$_2$Se$_{2-x}$O (x = 0, 0.5) are consistent with the tetragonal structure ($P4/mmm$) with the lattice parameters of $a$ = 3.979(8) ${\mathring{A}}$, $c$ = 7.934(2) ${\mathring{A}}$ for CsV$_2$Se$_2$O, and $a$ = 3.991(9) ${\mathring{A}}$, $c$ = 7.898(8) ${\mathring{A}}$ for CsV$_2$Se$_{1.5}$O. As we can see, with decreasing the selenium content, the lattice constant $c$ of this structure shrinks. After reacted with iodine, the Cs-ions are extracted from the structure. Due to extracting  Cs from the material, the structural stability requires the sliding of every neighbouring V$_2$Se$_2$O layers by 1/2 lattice constant along $a$- and $b$-axes. Hence, the space group changes from $P4/mmm$ in CsV$_2$Se$_2$O to $I4/mmm$ in V$_2$Se$_2$O. As shown in Figure~\ref{fig2}(c), the data can be well indexed with a tetragonal body-centered unit cell($I4/mmm$, $a$ = 3.841(3) ${\mathring{A}}$, $c$ = 11.658(4)${\mathring{A}}$), except for one or two very tiny peaks which are from impurity. As we can see, the peaks in Figure~\ref{fig2}(c) are broad for the sample V$_2$Se$_2$O. We address this broadening of peaks in the following way. Since the sample V$_2$Se$_2$O is obtained by deintercalating the Cs from the parent phase CsV$_2$Se$_2$O, the process may lead to the imperfect crystallinity of the sample. There are some coherently scattering domains which contribute to the x-ray diffraction peaks. For the Rietveld calculation, the width at the half-maximum of the peaks change with the average size of these domains. We did the simulation with different average domain sizes and show the results with the pink and green lines in Figure~\ref{fig2}(c). It can be clearly seen that the smaller domain size will lead to broader peaks. The calculation result gives the domain size of 12.9 nm for the V$_2$Se$_2$O polycrystal. The broad peaks of XRD patterns are also observed in Na$_{1-y}$Fe$_{2-x}$As$_2$\cite{NaFeAs} and tetragonal cobalt chalcogenides\cite{CoSeCoS} systems, which are also synthesized by the deintercalation method.

\begin{table}
  \caption{Crystallographic data of CsV$_2$Se$_2$O, CsV$_2$Se$_{1.5}$O and V$_2$Se$_2$O at 300 K.}
  \label{tbl}
  \begin{tabular}{lllllll}
    \hline
    \multicolumn{3}{l}{compound}  & \multicolumn{3}{l}{CsV$_2$Se$_2$O}  \\
    \hline
    \multicolumn{3}{l}{space group    }          & \multicolumn{3}{l}{$P4/mmm$}   \\
    \multicolumn{3}{l}{$a$ ($\mathring{A}$) }    & \multicolumn{3}{l}{3.979(8)}  \\
    \multicolumn{3}{l}{$c$ ($\mathring{A}$) }    & \multicolumn{3}{l}{7.934(2)}  \\
    \multicolumn{3}{l}{$V$ ($\mathring{A}^3$)  }   & \multicolumn{3}{l}{125.671(7)}  \\
    \multicolumn{3}{l}{$\rho$ (g/cm$^3$)  }   & \multicolumn{3}{l}{5.708}  \\
    \multicolumn{3}{l}{$R_{wp}$ (\%) }           & \multicolumn{3}{l}{7.54} \\
    \multicolumn{3}{l}{$R_{p}$ (\%) }           & \multicolumn{3}{l}{5.51} \\
    \multicolumn{3}{l}{$GOF$}           & \multicolumn{3}{l}{1.41} \\
    \hline
    atom & site & x & y & z & occupancy & B$_{eq}$ \\
    \hline
    Cs  & 1b & 0 & 0 & 0.5 & 1.09(8) & 0.2137 \\
    V & 2f & 0.5 & 0 & 0 & 1 & 0.5032 \\
    Se & 2h & 0.5 & 0.5 & 0.213(9) & 1.06(5) & 0.4607 \\
    O & 1a & 0 & 0 & 0 & 1 & 0.1059 \\
    \hline
    \hline
    \multicolumn{3}{l}{compound}  & \multicolumn{3}{l}{CsV$_2$Se$_{1.5}$O}  \\
    \hline
    \multicolumn{3}{l}{space group    }          & \multicolumn{3}{l}{$P4/mmm$}   \\
    \multicolumn{3}{l}{$a$ ($\mathring{A}$) }    & \multicolumn{3}{l}{3.991(9)}  \\
    \multicolumn{3}{l}{$c$ ($\mathring{A}$) }    & \multicolumn{3}{l}{7.898(8)}  \\
    \multicolumn{3}{l}{$V$ ($\mathring{A}^3$)  }   & \multicolumn{3}{l}{125.874(5)}  \\
    \multicolumn{3}{l}{$\rho$ (g/cm$^3$)  }   & \multicolumn{3}{l}{4.849}  \\
    \multicolumn{3}{l}{$R_{wp}$ (\%) }           & \multicolumn{3}{l}{7.00} \\
    \multicolumn{3}{l}{$R_{p}$ (\%) }           & \multicolumn{3}{l}{5.27} \\
    \multicolumn{3}{l}{$GOF$}           & \multicolumn{3}{l}{1.81} \\
    \hline
    atom & site & x & y & z & occupancy & B$_{eq}$ \\
    \hline
    Cs  & 1b & 0 & 0 & 0.5 & 0.93(6) & 0.6537 \\
    V & 2f & 0.5 & 0 & 0 & 1 & 0.5706 \\
    Se & 2h & 0.5 & 0.5 & 0.215(1) & 0.79(3) & 0.1018 \\
    O & 1a & 0 & 0 & 0 & 1 & 0.3942 \\
    \hline
    \hline
    \multicolumn{3}{l}{compound}  & \multicolumn{3}{l}{V$_2$Se$_2$O}  \\
    \hline
    \multicolumn{3}{l}{space group    }          & \multicolumn{3}{l}{$I4/mmm$}   \\
    \multicolumn{3}{l}{$a$ ($\mathring{A}$) }    & \multicolumn{3}{l}{3.841(3)}  \\
    \multicolumn{3}{l}{$c$ ($\mathring{A}$) }    & \multicolumn{3}{l}{11.658(4)}  \\
    \multicolumn{3}{l}{$V$ ($\mathring{A}^3$)  }   & \multicolumn{3}{l}{172.035(0)}  \\
    \multicolumn{3}{l}{$\rho$ (g/cm$^3$)  }   & \multicolumn{3}{l}{5.324}  \\
    \multicolumn{3}{l}{$R_{wp}$ (\%) }           & \multicolumn{3}{l}{3.47} \\
    \multicolumn{3}{l}{$R_{p}$ (\%) }           & \multicolumn{3}{l}{2.50} \\
    \multicolumn{3}{l}{$GOF$}           & \multicolumn{3}{l}{1.44} \\
    \hline
    atom & site & x & y & z & occupancy & B$_{eq}$ \\
    \hline
    V & 4c & 0.5 & 0 & 0 & 1 & 0.2339 \\
    Se & 4e & 0.5 & 0.5 & 0.368(7) & 1 & 0.8990 \\
    O & 2a & 0 & 0 & 0 & 1 & 0.1542 \\
    \hline
  \end{tabular}
\end{table}

\subsection{Transport and magnetic properties}

\begin{figure}
  \includegraphics[width=8.5cm]{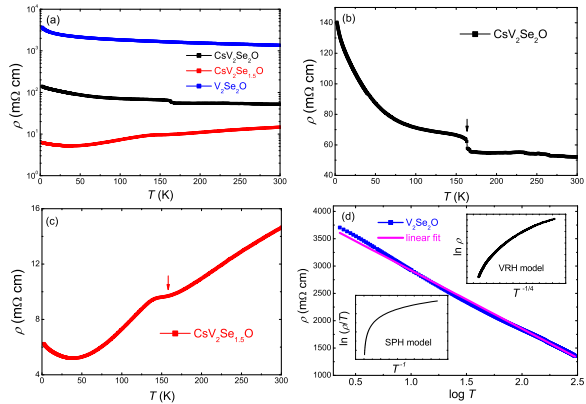}
\caption{(a) Temperature dependence of resistivity in semi-logarithmic scale for CsV$_2$Se$_2$O, CsV$_2$Se$_{1.5}$O and V$_2$Se$_2$O samples from 2 K to 300 K at ambient pressure and zero magnetic field. (b),(c) Temperature dependence of resistivity in linear coordinate for CsV$_2$Se$_2$O and CsV$_2$Se$_{1.5}$O, respectively. The arrows indicate the transitions for the two samples. (d) Temperature dependence of resistivity in the $\rho$ versus $log(T)$ for V$_2$Se$_2$O in the temperature range from 2 K to 300 K, together with the corresponding linear fit (pink line). The upper inset shows the curve of ln$\rho$ versus $T$$^{-1/4}$, where the linear case would correspond to the VRH model. And the lower inset shows the curve of ln$\rho$ versus $T$$^{-1}$, where the linear case would correspond to SPH model.} \label{fig3}
\end{figure}

Figure~\ref{fig3}(a) shows the comparison of the temperature dependence of resistivity in semi-logarithmic scale for CsV$_2$Se$_2$O, CsV$_2$Se$_{1.5}$O and V$_2$Se$_2$O samples from 2 K to 300 K at ambient pressure. It can be clearly seen that resistivity of these three samples have a considerable difference in magnitude and behavior. Details are shown in Figure~\ref{fig3}(b)$\sim$(d). The resistivity of CsV$_2$Se$_2$O increases while cooling and then reaches 140 m$\Omega$$\cdot$cm at 2 K. Thus, CsV$_2$Se$_2$O can be regarded as a semiconductor or a weak insulator. While in sharp contrast, the CsV$_2$Se$_{1.5}$O sample has a much smaller resistivity and shows a metallic behavior above 38 K with a slight upturn at low temperatures. The increase of resistivity with cooling below 38 K may be due to the weak localization effect or some other effects. For both compounds, an resistivity anomaly as marked by the arrows can be observed in the intermediate temperature region, which may correspond to the charge- or spin-density-wave(CDW/SDW) or structural transitions. The transition occurs at about 168 K in CsV$_2$Se$_2$O and at about 150 K in CsV$_2$Se$_{1.5}$O. This CDW/SDW-like transition was also observed in other $A$${T}$$_2$$Q$$_2$O compounds, such as BaTi$_2$Sb$_2$O($T_s$ = 50 K)\cite{BaTiSbO} and BaTi$_2$As$_2$O($T_s$ = 200 K)\cite{BaTiAsO}. The intrinsic reason for these transitions is still unknown, and more experiments should be done to identify them. In contrast, this kind of transition was not seen in CsV$_2$S$_2$O.\cite{CsVSO}

However, the V$_2$Se$_2$O sample exhibits a very distinct electric transport property, as shown in Figure~\ref{fig3}(a) and (d). Firstly, the resistivity of V$_2$Se$_2$O is much larger than that of CsV$_2$Se$_2$O and CsV$_2$Se$_{1.5}$O. Secondly, with decreasing temperature, it increases monotonously without any anomaly. Surprisingly, the curve of resistivity versus log$T$ becomes a nearly-straight line with a negative slope from 2 K to 300 K. In other words, we find the $\rho\propto$log$(1/T)$ in wide temperature range below 300 K. However, neither its parent compound CsV$_2$Se$_2$O nor CsV$_2$Se$_{1.5}$O has shown this kind of behavior. We also try to fit the temperature dependence of resistivity with other functions, such as the thermal activation model for a band insulator (ln$\rho \propto 1/T$), or the three-dimensional variable-range-hopping(VRH) model (ln$\rho \propto T^{-1/4}$, shown in the upper-right inset), and small-polaron-hopping(SPH) model(ln($\rho/T) \propto -1/T$, shown in the bottom-left inset), but all these models fail to fit the data. This implies that the V$_2$Se$_2$O system is an insulator with an unusual electric transport mechanism. This kind of logarithmic behavior, namely $\rho\propto$log$(1/T)$ was actually reported in the parent phase or normal state of some cuprates, like underdoped La$_{2-x}$Sr$_{x}$CuO$_4$\cite{LaSrCuO} (at 60 T) and overdoped Bi$_2$Sr$_{2-x}$La$_{x}$CuO$_{6+\delta}$\cite{BiSrLaCuO}. In these systems, the logarithmic behaviors are all observed at low temperatures. But in the V$_2$Se$_2$O system, the logarithmic behavior is so robust and exhibits in a wide range of temperature. At the moment we cannot give the explicit reasons for this behavior. However we believe that it is related to the electron correlation effect, and probably it is a consequence of a Mott insulator.

\begin{figure}
  \includegraphics[width=8.5cm]{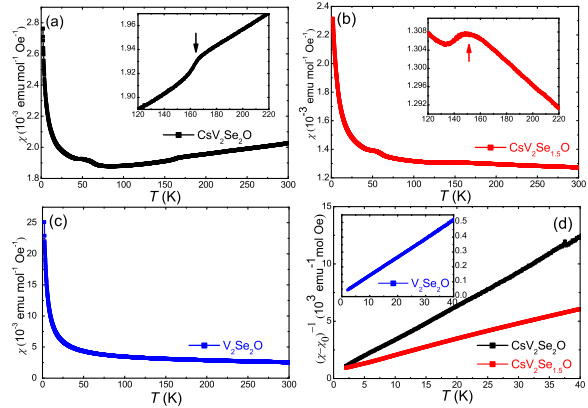}
\caption{Temperature dependence of magnetic susceptibility measured at 1 T for (a) CsV$_2$Se$_2$O, (b) CsV$_2$Se$_{1.5}$O and (c) V$_2$Se$_2$O polycrystalline samples. The insets of (a) and (b) show clearly two small kinks at 168 K for CsV$_2$Se$_2$O and 150 K for CsV$_2$Se$_{1.5}$O, at these temperatures we find anomalies in their $\rho(T)$ curves. (d) Temperature dependence of 1/($\chi$-$\chi_0$) for CsV$_2$Se$_2$O, CsV$_2$Se$_{1.5}$O and V$_2$Se$_2$O from 2 K to 40 K. One can obtain a linear relation between ($\chi$-$\chi_0$)$^{-1}$ and $T$ in the low temperature region.} \label{fig4}
\end{figure}

Figures~\ref{fig4}(a), (b) and (c) shows the temperature dependence of magnetic susceptibility for CsV$_2$Se$_2$O, CsV$_2$Se$_{1.5}$O and V$_2$Se$_2$O polycrystalline samples, respectively. They are all measured at a magnetic field of 1 T with zero-field-cooling mode. With decreasing temperature, the magnetic susceptibility of CsV$_2$Se$_2$O decreases slowly from 300 K to 75 K, and then increases rapidly below 75 K. But the magnetic susceptibility of CsV$_2$Se$_{1.5}$O is always increasing with cooling down. Besides, it is worth to mention that two small kinks can be seen at 168 K for CsV$_2$Se$_2$O and 150 K for CsV$_2$Se$_{1.5}$O as shown in the insets. At the same temperatures, two anomalies are also found in their resistivity curves in Figure~\ref{fig3}. However, the magnetic susceptibility of V$_2$Se$_2$O increases while cooling without any kinks, just as the performance in the $\rho(T)$ curve. We have not found any trace of ferromagnetic order in all these samples. Furthermore, the temperature dependence of magnetic susceptibility of all samples show a divergence in low temperature region.

To further analyze the average magnetic moments from the V-atoms in each sample, we try to use the Curie-Weiss law to fit the data in low temperature region,
\begin{equation}
  \chi = M/H = \chi_0 + \frac{C}{T + T_\theta}.
\end{equation}
where $\chi_0$, $T_\theta$, and $C$ are the fitting parameters. The $\chi_0$ comes from the Pauli paramagnetism of the conduction electrons, which is related to the density of states (DOS) at the Fermi energy. $C$ = $\mu_0$$\mu_{eff}^2$/3$k_B$, where $\mu_{eff}$ is the local magnetic moment per V-site. By adjusting $\chi_0$, this formula can also be written in a linear form, ($\chi$-$\chi$$_0$)$^{-1}$ = $T$/$C$ + $T_\theta$/$C$. The slope of ($\chi$-$\chi_0$)$^{-1}$ versus $T$ gives the value of 1/$C$ and the intercept gives $T_{\theta}$/$C$. Figure~\ref{fig4}(d) presents the temperature dependence of ($\chi$-$\chi_0$)$^{-1}$ from 2 K to 40 K for CsV$_2$Se$_2$O ($\chi_0$ = 1.85024$\times$10$^{-3}$ emu/mol$\cdot$Oe), CsV$_2$Se$_{1.5}$O ($\chi_0$ = 1.24256$\times$10$^{-3}$ emu/mol$\cdot$Oe) and V$_2$Se$_2$O ($\chi_0$ = 2.8075$\times$10$^{-3}$ emu/mol$\cdot$Oe). The obvious linearity indicates that the data in the low temperature region can be fitted by the Curie-Weiss law perfectly. From the slopes of these lines, we obtain the magnetic moments per V-site $\mu_{eff}$ = 0.117$\mu_B$ for CsV$_2$Se$_2$O, $\mu_{eff}$ = 0.169$\mu_B$ for CsV$_2$Se$_{1.5}$O and $\mu_{eff}$ = 0.569$\mu_B$ for V$_2$Se$_2$O. Clearly the V$_2$Se$_2$O sample has the largest magnetic moment. Combining with the resistivity data, it is reasonable to believe that the V$_2$Se$_2$O system with a larger magnetic moment has a stronger electronic correlation which leads to a worse electronic conduction.

\section{Conclusions}
In conclusion, we successfully synthesize several new layered vanadium based compounds, CsV$_2$Se$_{2-x}$O(x = 0, 0.5) and V$_2$Se$_2$O. By changing the deficiency of Se, or completely removing the Cs we can tune the valence state of vanadium, and the system can be tuned from a metal (CsV$_2$Se$_{1.5}$O) to a strongly correlated insulator (V$_2$Se$_{2}$O). In the system V$_2$Se$_{2}$O we find the $\rho\propto$log$(1/T)$ relation which is similar to that discovered in the parent phase or the ground state of some cuprate superconductors. We attribute this feature to the strong correlation effect. The vanadium based systems found here may provide very good platforms for investigating the balance between the localization and itinerancy of 3$d$ electrons.

\section*{ACKNOWLEDGMENTS}
This work was supported by the Ministry of Science and Technology of China (Grant No. 2016YFA0300401, 2016YFA0401704), and the National Science Foundation China (NSFC) with the projects: A0402/11534005, A0402/11674164.


\begin{thebibliography}{00}
\bibitem{LaBaCuO}J. G. Bednorz, and K. A. Z. M$\ddot{u}$ller, Phys. B \textbf{64}, 189-193 (1986).
\bibitem{LaOFFeAs}Y. Kamihara, T. Watanabe, M. Hirano, and H. Hosono, J. Am. Chem. Soc. \textbf{130}, 3296-3297 (2008).
\bibitem{BaTiSbO}T. Yajima, K. Nakano, F. Takeiri, T. Ono, Y. Hosokoshi, Y. Matsushita, J. Hester, Y. Kobayashi, and H. Kageyama, J. Phys. Soc. Jpn. \textbf{81}, 103706 (2012).
\bibitem{BaNaTiSbO}P. Doan, M. Gooch, Z. J. Tang, B. Lorenz, A. M$\ddot{o}$ller, J. Tapp, P. C. W. Chu, and A. M. Guloy, J. Am. Chem. Soc. \textbf{134}, 16520-16523 (2012).
\bibitem{BaTiBiO}T. Yajima, K. Nakano, F. Takeiri, J. Hester, T. Yamamoto, Y. Kobayashi, N. Tsuji, J. Kim, A. Fujiwara, and H. Kageyama, J. Phys. Soc. Jpn. \textbf{82}, 013703 (2013).
\bibitem{LaOCoSe}C. Wang, M. Q. Tan, C. M. Feng, Z. F. Ma, S. Jiang, Z. A. Xu, G. H. Cao, K. Matsubayashi, and Y. Uwatoko, J. Am. Chem. Soc. \textbf{132}, 7069-7073 (2010).
\bibitem{LaOFeSe}J. M. Mayer, L. F. Schneemeyer, T. Siegrist, J. V. Waszczak, and B. V. Dover, Angew. Chem. Int. Ed. Engl. \textbf{31}, 1645 (1992).
\bibitem{LaONiSe}C. Le, J. Zeng, G. H. Cao, and J. P. Hu, arXiv:1712.05962v1 (2017).
\bibitem{CsVSO}M. Valldor, P. Merz, Y. Prots, and W. Schnelle, Eur. J. Inorg. Chem. 23-27 (2016).
\bibitem{NaFeAs}G. M. Friederichs, I. Schellenberg, R. P$\ddot{o}$ttgen, V. Duppel, L. Kienle, J. S. auf der G$\ddot{u}$nne, and D. Johrendt, Inorg. Chem. \textbf{51}, 8161-8167 (2012).
\bibitem{MoS}Y. Q. Fang, J. Pan, J. Q. He, R. C. Luo, D. Wang, X. L. Che, K. J. Bu, W. Zhao, P. Liu, G. Mu, H. Zhang, T. Q. Lin, and F. Q. Huang, Angew. Chem. Int. Ed. \textbf{57}, 1232-1235 (2018).
\bibitem{LiMoS}C. G. Guo, H. Li, W. Zhao, J. Pan, T. Q. Lin, J. J. Xu, M. W. Chen, and F. Q. Huang, J. Mater. Chem. C \textbf{5}, 5977 (2017).
\bibitem{FeSe}D. N. Yuan, Y. L. Huang, S. L. Ni, H. X. Zhou, Y. Y. Mao, W. Hu, J. Yuan, K. Jin, G. M. Zhang, X. L. Dong, and F. Zhou, Chin. Phys. B \textbf{25}, 077404 (2016).
\bibitem{FeS}H. Lin, Y. F. Li, Q. Deng, J. Xing, J. Z. Liu, X. Y. Zhu, H. Yang, and H. H. Wen, Phys. Rev. B \textbf{93}, 144505 (2016).
\bibitem{CoSeCoS}X. Q. Zhou, B. Wilfong, H. Vivanco, J. Paglione, C. M. Brown, and E. E. Rodriguez, J. Am. Chem. Soc. \textbf{138}, 16432-16442 (2016).
\bibitem{Rietveld}H. M. Rietveld, J. Appl. Crystallogr. \textbf{2}, 65-71 (1969).
\bibitem{TOPAS}R. W. Cheary, and A. Coelho, J. Appl. Crystallogr. \textbf{25}, 109-121 (1992).
\bibitem{BaTiAsO}X. F. Wang, Y. J. Yan, J. J. Ying, Q. J. Li, M. Zhang, N. Xu, and X. H. Chen, J. Phys.: Condens. Matter \textbf{22}, 075702 (2010).
\bibitem{LaSrCuO}Y. Ando, G. S. Boebinger, A. Passner, T. Kimura, and K. Kishio, Phys. Rev. Lett. \textbf{75}, 4662 (1995).
\bibitem{BiSrLaCuO}S. Ono, Y. Ando, T. Murayama, F. F. Balakirev, J. B. Betts, and G. S. Boebinger, Phys. Rev. Lett. \textbf{85}, 638 (2000).

\end{thebibliography}
\end{document}